\documentclass[granma,final]{svjour}
\usepackage{graphicx}
\usepackage{times}
\begin{document}
\title{
{\it Temperature}
measurement under the convection and segregation in the vibrated
bed of
powder:
A numerical study
}
\author{Satoru Kiyono and Y-h. Taguchi% etc
% The present address - made with \thanks - is optional,
% remove next line if not needed
}                     % Do not remove
\mail{Y-h. Taguchi}
\institute{Department of Physics, and  Institute for
Science and Technology, Chuo University, Tokyo 112-8551
\email{tag@granular.com}}
\date{Received: date / Revised version: date}
\maketitle
\begin{abstract}
In  numerically simulated vibrated beds of powder,
we measure temperature under  convection
by the generalized Einstein's relation.
The spatial temperature distribution  turns out to be 
quite uniform except for the boundary layers.
In addition to this, temperature remains uniform even if
segregation occurs. 
This suggests the possibility that
there exists some {\it thermal equilibrium state} even in a
vibrated bed of powder. This finding may lead to a
unified view of the dynamic steady state of granular matter.
\end{abstract}

\section{Introduction}
%\label{intro}
Granular matter is characterized by a set of macroscopic
particles, e.g., sand, snow fall, sugar, salt and rice.
Recently many physicists became interested in the dynamics of
granular matter \cite{RMP}, because it shows many striking features.
These include convection, segregation, surface waves, and
flows. They are hard to understand by the word of physics,
although it is relatively easy to reproduce them numerically.

For example, the concept of granular temperature \cite{RMP} is often used
to describe the dynamical state of powder.
However, in contrast to  rapid granular flow \cite{Goldhirsch}, 
the granular temperature turns out not to be a good measure
for dense granular matter.
For example, granular temperature in vibrated beds of powder
is anisotropic \cite{gra_temp},
which is not a favorable feature.

On the other hand, Wildman and Huntley \cite{wildman} have found useful a
temperature that is given by the mean squared displacement curves for short
times. Although it sounds promising, their definition is valid only when
convective flow is negligible.

Recently Makse and Kurchan \cite{makse}
have numerically found that the temperature defined by Einstein's 
relation is well defined  and is independent of the size of
the particles in granular matter.
In contrast to that of Wildman and Huntley \cite{wildman}, 
Makse and Kurchan's
definition is valid even if there is a non-negligible flow.
Also D'Anna {\it et al} \cite{D'Anna} 
have found that
their variant using AC frequency is also well defined.
Thus we may expect that transport phenomena of granular particles
can give us a more suitable definition of temperature than
the conventional granular temperature.

In spite of these successful applications of
Einstein's relation to define the temperature in the granular matter,
its application is very limited.
This is because the conventional Einstein's relation  can be
defined only in a spatially uniform system.
In order to apply Einstein's relation, we need to understand the
long time behavior of tagged particles.
If the system is not uniform at all, the tagged particle
will travel around  various places having various temperatures.
Thus what we can measure is a spatially averaged temperature, which
cannot reflect spatial structure in the  system.
On the other hand, if the system is uniform,  temperature is
uniform, too. This means that we can never 
know if the thermal equilibrium really stands or not
from the measurement of temperature by Einstein's relation.
However, as demonstrated by Wildman and  Huntley \cite{wildman},
it is not always necessary to observe long time behavior in order to
derive temperature from the transport phenomena.

In this paper, we have developed a heuristic method to
measure temperature by Einstein's relation in the strongly non-uniform
system. In \S \ref{sec1}, following Makse and Kurchan \cite{makse},
we introduce the temperature using Einstein's  
relation. 
In \S \ref{sec2}, the heuristic approach to define local
temperature is presented, and in \S \ref{sec3} this definition is further
generalized so as to measure temperature in the system under the flow.
The numerical results in vibrated beds are shown in \S \ref{sec4},
and how  segregation affects the temperature is investigated in
\S \ref{sec5}. 
In \S \ref{sec6}, we have discussed about the validity of our results;
\S \ref{sec7} contains summary and conclusion.

\section{Einstein's relation and temperature}
\label{sec1}
%%%%%%%

Suppose there is a particle which obeys a  random walk.
So as to be simple, we restrict its motion to the one dimensional space.
If $x(t)$ is  the coordinate of the particle at time $t$, then
the diffusion constant $D$ is defined as
$$
\langle [x(t) - x(0)]^2 \rangle =2Dt,
$$
where $\langle \cdots \rangle$ means the ensemble average.
On the other hand, the mobility $\mu$ can be defined  as
$$
\langle x(t)-x(0)\rangle = \mu f t,
$$
where $f$ is the applied force.
When combining the above two equations with the Einstein's relation
$$
T \equiv \frac{D}{\mu},
$$
where $T$ is temperature,
we get
$$
 T = \frac {f}{2} 
\frac{\langle [x(t) - x(0) ]^2\rangle }
{\langle x(t) -x(0)\rangle}.
$$

Recently, Makse and Kurchan \cite{makse} used this equation
as the definition of temperature in granular matter.
They have performed three dimensional distinct element method (DEM) for the
periodic sheared system consisting of a binary set of
both large and small particles.
They measured $D$ and $\mu$ along the direction perpendicular to the
shear and found that the measured temperature is the same for
both large and small particles.

More recently, D'Anna {\it et al} \cite{D'Anna} have experimentally found that
the fluctuation-dissipation  ratio
$$
\frac{S(\omega)\omega}{4 \chi''(\omega)}
$$
can be a well defined temperature for the vibrated bed,
where $S(\omega)$ is the noise power spectrum density  of the 
angular frequency $\omega$ and 
$\chi''$ is the imaginary part of the complex susceptibility.
Since this can be regarded as the generalization of the temperature
introduced by Makse and Kurchan to the frequency dependent one,
these kinds of definition of temperature seem to be valid for
granular matter.

However, there is a difficulty to apply the above definitions to
general cases. This is because it is impossible to
measure the spatial dependence of temperature using these definitions.
Makse and Kurchan's definition is valid only for the long time limit, and
D'Anna {\it et al}'s procedure is used to measure temperature
only when we can regard a whole vibrated bed as a thermal bath.
Thus, it is rather difficult to relate temperature to the
statistical mechanics, because in the statistical mechanics
local temperatures must be defined. 
The definitions above cannot provide  such information.

\section{Heuristic definition of local temperature for a
one dimensional system}\label{sec2}

In order to generalize Makse and Kurchan's definitions
to measure local temperature,
we propose a  heuristic procedure.
First, we subdivide the whole system into smaller cells.
Each cell has an index $i$. 
Next, we define the transition probability of a particle from the $i$th cell
to the $j$th cell $P(j,i)$. Suppose the spacing between cells is $\ell$,
a random walker must travel the distance of $\ell$ in order to go
from cell $i$ to cell $i+1$.
Thus the averaged time $t_0$  until this arises is obtained as
$$
t_0 = \ell^2/ 2D.
$$
The  escape probability of
a particle from  cell $i$  is
$$
P_{out} (i) \equiv \sum_{k=\pm 1} P(i+k,i; f=0) \propto 1/t_0 = 2D / \ell^2.
$$
On the other hand, when a drift force $f$ exists,
$$
t_0 = \ell / f\mu,
$$
then the flow $J(i)$ along the $f$ direction is
$$
J(i) \equiv \ell \sum_{k=\pm 1} kP(i+k,i; f\neq 0) \propto \ell/t_0 = f \mu.
$$
Thus  
$$
 T_i = \frac{D}{ \mu} = 
\frac{ P_{out}(i) \ell^2/2}{J(i)/f} =
\frac{f}{2} \frac{P_{out}(i) \ell^2}{J(i)}
$$
is the  local temperature at  cell $i$.
Of course we do not insist that this is rigorous,
but only regard it as a heuristic argument.
Also, in order to remove the dependency on  $f$
we employ the following definition 
$$
 T^{\mbox{eff}}_i \equiv \lim_{f \rightarrow 0} T_i
$$
of the effective temperature.

\section{The definition of effective temperature in the system with
flow}
\label{sec3}

As an example to test the above heuristic definition of the
effective temperature,
we employ a numerical simulation of a vibrated bed of powder.
In the vibrated bed of powder considered here, 
there are convective motions of particles.
Thus $\langle x(t)-x(0) \rangle$ is not zero even if $f$ is equal to $0$.
Furthermore, the vibrated bed we deal with is a two dimensional system.
Therefore we need some additional 
modification of the previous definition of the
effective temperature $T^{\mbox{eff}}_i$.

\begin{figure}[h]
\begin{center}
\includegraphics[scale=0.3]{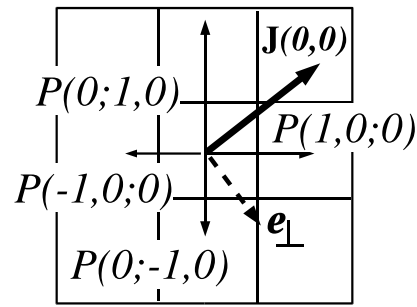}
\caption{ 
Schematic figure to explain the relationship among
the transition probability $P$, the flow vector {\boldmath $J$} and 
the unit vector $\mbox{\boldmath $e$}_\perp$
perpendicular to the flow vector.
In this example, the center cell is numbered as $(0,0)$.
Thin solid arrows indicate the transition probability,
bold solid arrow indicates the flow vector, and
broken arrow indicates 
the unit vector $\mbox{\boldmath $e$}_\perp$
perpendicular to the flow vector, 
along which the force is applied later.
 The flow vector {\boldmath $J$} of the center cell is
$\ell ( P(1,0;0) -P(-1,0;0),P(0;1,0) - P(0;-1,0))$.
The escape  probability of a particle from the center cell
$\mbox{\boldmath $P$}^{out}$ is $( P(1,0;0) +P(-1,0;0),P(0;1,0) +P(0:-1,0))$.}
\label{fig:schem}
\end{center}
\end{figure}

First we subdivide the system into $N_1 \times N_2$ cells,
where $N_1 \ell$ is the horizontal size of system and $N_2 \ell$ 
is the vertical size of the system. Then we denote the cell as $(i,j)$  if the
cell is the $i$th in horizontal direction and the $j$th in vertical direction.
The flow vector at  cell $(i,j)$ is defined as
$$
 {\mbox{\boldmath $J$}}(i,j; f=0)  \\ 
 \equiv \ell \sum_{k=\pm 1}  k \left ( 
\begin{array}{c}
P(i+k,i;j ; f=0) \\
P(i; j+k,j ;f=0)
\end{array} 
 \right),
$$
where $P(i',i;j)$ is the transition probability from
$(i,j)$ to $(i',j)$ and
 $P(i;j',j)$ is the transition probability from
$(i,j)$ to $(i,j')$ (See Fig. \ref{fig:schem}).

%\hspace{7.5cm}  \fbox{Fig.  \ref{fig:schem}}

Next we can define the escape probability
of a particle from the cell $(i,j)$ as
$$
 P_x^{out} (i,j; f=0)  \\ 
 \equiv  \sum_{k=\pm 1}  P(i+k,i;j ; f=0).
$$
Similarly,
$$
 P_y^{out} (i,j; f=0)  \\ 
 \equiv  \sum_{k=\pm 1}  P(i;j+k,j ; f=0).
$$

In order to deal with vibrated beds, we have to take into consideration the case
when {\boldmath $J$} is not zero.
We have to find the direction along which the flow is zero when $f$
is equal to zero. 
Such a direction is perpendicular to the flow vector.
Thus the unit vector along this direction is
$$
\mbox{\boldmath$e$}_\perp(i,j;f=0) = 
\frac{1}{ | \mbox{\boldmath $J$}|}
\left( 
\begin{array}{c}
-J_y(i,j;f=0) \\
J_x(i,j;f=0) 
\end{array}
\right).
$$
The escape probability of a particle from the $i$th cell along this
direction is defined as
$$
P_\perp^{out} (i,j ;f=0)  \equiv | \mbox{ \boldmath$e$}_\perp(i,j;f=0) 
\cdot  \mbox{\boldmath$P$}^{out}(i,j; f=0) |,
%\left(  e_x(i,j) J_x(i,j;f=0) \right . \\
%& +  & \left .e_y (i,j) J_y(i,j ; f=0) \right) / \ell.
$$
where 
$$
\mbox{\boldmath $P$}^{out}(i,j;f=0) =
\left( 
\begin{array}{c}
P_x^{out}(i,j;f=0)\\
P_y^{out}(i,j;f=0)
\end{array}
\right).
$$

Next we apply a small force $f \mbox{\boldmath $e$}_\perp$ at  cell $(i,j)$.
This time, the flow along this axis is defined as
$$
J_\perp (i,j ;f\neq0)  \equiv 
 \mbox{\boldmath $e$}_\perp(i,j;f=0) 
\cdot  \mbox{\boldmath $J$}(i,j; f \neq 0) ,
%e_x(i,j) J_x(i,j;f \neq 0) \\
%& +  &  e_y (i,j) J_y (i,j;f \neq 0) .
$$
which takes non-zero values when $f \neq 0$.
Using these above, we get
$$
 T_{ij} = \frac{f}{2} 
\frac{ P_\perp^{out}(i,j; f = 0) \ell^2}
{ J_\perp(i,j; f\neq 0)},
$$
as a local temperature at $(i,j)$.
$T^{\mbox{eff}}_{ij}$ is obtained after taking the limit of $f \rightarrow 0$
as before.

\section{Numerical measurement of the effective granular temperature}
\label{sec4}

\begin{figure}[h]
\begin{center}
\includegraphics[scale=0.5]{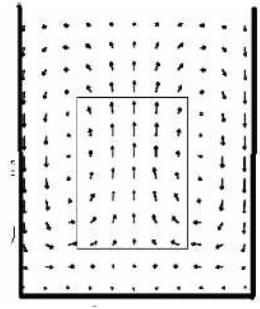}
\caption{The flow vectors {\boldmath $J$}$(i,j)$. 
The flow is upward at the center region and
downward at the wall region. The solid rectangle
($ 4 \leq i \leq 8, 3 \leq j \leq 9$) shows the region
where the effective temperature will be computed.
The left-bottom corner corresponds to $(i,j)=(1,1)$ and
$i(j)$ increases right(up)wards to $i_{\mbox{max}} = 11$ and $j_{\mbox{max}} =13$. }
\label{fig:conv}
\end{center}
\end{figure}

The DEM method is  the same as in the previous study\cite{tag}.
The number of particles is 100,
the diameter of a particle is 1, and the horizontal size of the vessel is 11.
The coefficient of restitution is $e=0.79$ and $t_c=0.04$, with $k=3000$ (spring constant) and
$\eta=6$ (dissipation constant).
The strength of interactions with the walls and the bottom is  two times  larger 
than  the above $k$ and $\eta$.
The acceleration amplitude $\Gamma
= 1.51g$ ($g$ is the 
gravitational acceleration, here it is taken to be 98,
amplitude is 4.11 and angular frequency is 6.20).
Figure \ref{fig:conv} shows the $\mbox{\boldmath $J$}(i,j)$ averaged over 
sequential $4 \times 10^8$ snapshots, 
where $\ell =1$. This means $N_x=11 (i=1,..,11)$.
$N_y$ can be taken as large as possible, but in the present study
 $j=1,{\ldots} ,N_y(= 10)$.
The $(1,1)$ cell corresponds to the down-left corner.
$i$ and $j$ increases rightwards and upwards, respectively.

%\hspace{7.5cm}  \fbox{Fig. \ref{fig:conv}}

As can be seen easily, there is a convective motion over the whole system.
The flow pattern is not uniform at all.
Also the gravity force is applied. This means that
in a naive sense, it is difficult to expect a  homogeneous state
over the whole system. Thus it will be interesting to see
whether thermal equilibrium is kept or not using the above defined
effective temperature.

\begin{figure}[h]
\begin{center}
\includegraphics[scale=0.8]{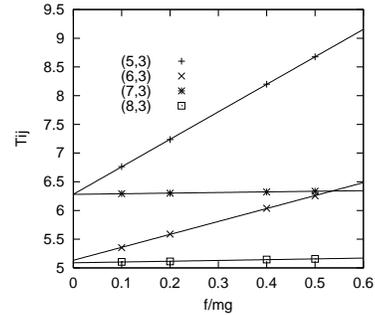}
\caption{The dependence of $ T_{ij}$ upon $f/mg$ at $j=3$.
Solid lines indicate a linear extrapolation to $f=0$ using
a least square fit.
The numbers in parentheses indicate the targeted cells as
$(i,j)$.}
\label{fig:T_i}
\end{center}
\end{figure}

In order to compute $ T^{\mbox{eff}}_{ij}$, first we 
compute $P(i \pm 1, i; j; f=0)$ and
$P(i; j\pm1, j; f=0)$ numerically. Then applying $f$, we  compute
$P(i \pm 1, i; j; f \neq  0)$ and $P(i; j\pm1, j; f \neq 0)$.
$f$ should be as small as possible, but too small $f$ cannot generate 
large enough $J_\perp(i,j;f \neq 0)$ to be estimated. 
Also if applying $f$ to all the cells
at once, global flow structures can be destroyed. Furthermore, we 
need measurements for several forces before 
taking $f \rightarrow 0$ limit.
Considering these requirements, we apply $f/mg= 0.1, 0.2,0.4,0.5$
to only a few cells at once ($m$ is the particle mass). 
Then repeat the same calculation after  changing 
the set of cells to which $f$ is applied.
Figure \ref{fig:T_i} shows the dependence of $ T_{ij}$ upon $f/mg$
at cells with $j=3$.
Since they have clear linear dependence upon $f$, we extrapolate
to $f=0$ using a linear least square fit in order to 
compute $T^{\mbox{eff}}_{ij}$.
Since the dependence of $T_{ij}$ upon $f/mg$ is similar for other layers,
we omit the remaining plots.

%\hspace{7.5cm}  \fbox{Fig. \ref{fig:T_i}}

\begin{figure}[h]
\begin{center}
\includegraphics[scale=0.5]{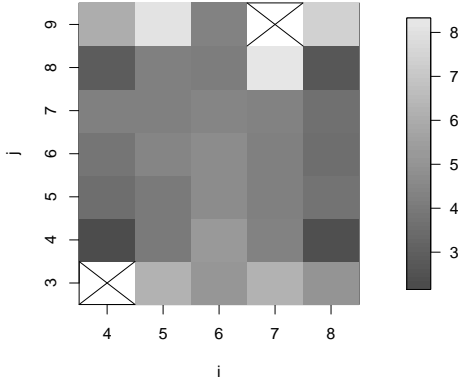}
\caption{Spatial distribution of $T^{\mbox{eff}}_{ij}$.
Each row corresponds to a horizontal layer of cells.
The whole region corresponds to the rectangle in
Fig. \ref{fig:conv}.
Missing values $(i,j) =(4,3),(7,9)$ are omitted due to the large error.}
\label{fig:T_ij}
\end{center}
\end{figure}

Figure \ref{fig:T_ij} shows the spatial distribution of $ T^{\mbox{eff}}_{ij}$.
We have omitted the regions close to the wall and the bottom ($4 \leq i \leq
8$, $ 3 \leq j \leq 9$).
Although  $ T^{\mbox{eff}}_{ij}$ fluctuates from cell to cell,
the values seem to be close to each other
(at least for $i=5,6,7$).
Thus surprisingly this suggests the system is more or less in
the {\it thermal equilibrium} state.

%\hspace{7.5cm}  \fbox{Fig. \ref{fig:T_ij}}

In order  to confirm this conclusion, we  propose the correction equation
of $T^{\mbox{eff}}_{ij}$. Why are these effective temperatures not constant?
Does  it mean the breakdown of thermal equilibrium?
As mentioned above,  $T_{ij}$ should be measured
along the axis without flow when $f=0$.
For this purpose, we selected the direction perpendicular to the flow
direction {\boldmath$J$}. However when measuring the effective temperature,
a particle can be  drifted to the neighboring cells where
the flow along drift direction is not zero.
In other words, in order that our formulation is satisfied well,
the flow in each cell has to
 be as  parallel to each other
 as it is in Makse and Kurchan's study. Of course, this requirement is
not satisfied in the  vibrated bed of powder at all.

\begin{figure}[h]
\begin{center}
\includegraphics[scale=0.5]{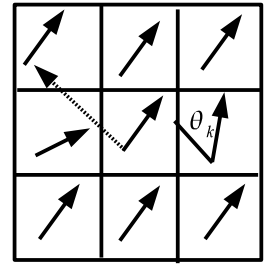}
\caption{Schematic figure to show the meaning of the correction equation
of the effective temperature $T^{\mbox{eff}}_{ij}$. Solid arrows indicate
the flow vectors at each cell.
When we compute the effective temperature at the center cell,
we apply a small force $f$  (dotted arrow) to the center cell.
However, if the flow vectors at surrounding cells  are not parallel to that of
the center cell, these accelerate or decelerate the drift caused by $f$.
This occurs when    
 the angle $\theta_k$ between the flow vector of surrounding cell and
the force direction differs from $\pi/2$.}
\label{fig:correct}
\end{center}
\end{figure}

\begin{figure}[h]
\begin{center}
\includegraphics[scale=0.8]{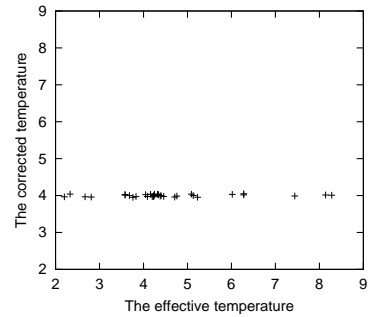}
\caption{
The corrected temperature as a function of the effective temperature
$ T^{\mbox{eff}}_{ij}$.
In spite of the large fluctuation of the effective temperature, the
corrected temperature takes an almost constant value.}
\label{fig:correct_T}
\end{center}
\end{figure}

In order to remove the effect that comes from the fact that the flow 
vectors are not parallel to each other, we propose here the heuristic equation,
$$
T^{\mbox{eff}}_{ij} =  \frac{T_0}{1 + \sqrt{2}}\left(1  + \sum_k \sin (\theta_k -
\frac{\pi}{2}) \right),
$$
where $T_0$ is the unperturbed true constant effective temperature, 
and the summation is taken over the eight neighboring cells, $k$.
$\theta_k$ is the angle between the flow vector and the direction of $f$ 
at cell $k$ (Fig. \ref{fig:correct}). 
The supposed meaning of this equation is
as follows. $\sin (\theta_k -\frac{\pi}{2})$ is the measure of the flow
along $f$ direction relative to neighboring cells. 
It is proportional to the difference between the neighboring flow vectors 
if we ignore the difference of absolute values of the vectors as higher order
corrections. Thus if it is negative(positive), particles flow 
in the same(opposite) direction of $f$.  
This apparently 
 increases(decreases) the drift caused by $f$, i.e. the denominator of the
definition of the effective temperature, 
therefore the measured temperature will be smaller
(larger) than the true temperature
$T_0$. In order to express this tendency we add $\sin (\theta_k -\frac{\pi}{2})$
to the right hand side.
$1 + \sqrt{2}$ is just a numeric parameter that
accounts for the square lattice.
The factor 1 represents distance between neighboring cells and
$\sqrt{2}$ does that between next neighboring cells.
It does not mean anything special at the present study.
Figure \ref{fig:correct_T} shows  
$$
T_0 = \frac{T^{\mbox{eff}}_{ij} ( 1 + \sqrt{2})}{\left[1  + \sum_k \sin (\theta_k -
\frac{\pi}{2}) \right] }.
$$
It takes an almost constant value for all cells.
Considering that there are no fitting parameters,
our heuristic argument turns out to be very good.
Thus we can conclude that the vibrated bed of powder is surely in
a thermal equilibrium state.

%\hspace{6.5cm}  \fbox{Fig. \ref{fig:correct}}
% \fbox{Fig. \ref{fig:correct_T}}

\section{Does segregation break the thermal equilibrium?}
\label{sec5}

In the previous section, the effective temperature
calculated from the heuristic argument indicates
that the vibrated bed of powder is in the thermal equilibrium state.
Although the kinetic property, i.e. flow, is non-uniform in the vibrated bed
of powder, particles are the same all over the vessel.
Thus it is interesting to see if  different particles can also be
in the thermal equilibrium state or not.
Actually, Makse and Kurchan \cite{makse} have shown
that particles with distinct sizes have the same effective temperature.
However,  their system is kinetically uniform.

In this section, we  deal with the vibrated bed of powder
with different particles. There are many ways to introduce differences
among particles. The size difference is the most famous
in the vibrated bed of powder, because it causes segregation.
However, in our cell approach, different sizes may cause some difficulty
because we have employed cells  as large as each particle.
Different particle size requires  different size of cells
or another approach.

Let us remind readers that  
the mixture of glass beads and lead beads causes segregation
\cite{ooyama,akiyama}. This phenomenon has already been  reproduced by
our model \cite{saito1,saito2} when we use two kinds of
particles with a distinct coefficient of restitution.
Thus the segregated state by the
introduction of the difference of the coefficient of restitution
will provide us the opportunity to see that the thermal equilibrium is
satisfied even if the system is spatially non-uniform in both
kinetics and material properties.
The coefficient of restitution used is 
$e=0.67 \; (t_c=0.04) \; [k=3200,\eta=10]$ and $0.88 \; (t_c=0.067) \; [k=1200,\eta=2]$.
The strength of interactions with walls and  bottom is  two times  larger.
The acceleration amplitude $\Gamma$ is taken to be $1.62g$.

\begin{figure}[h]
\begin{center}
\includegraphics[scale=0.5]{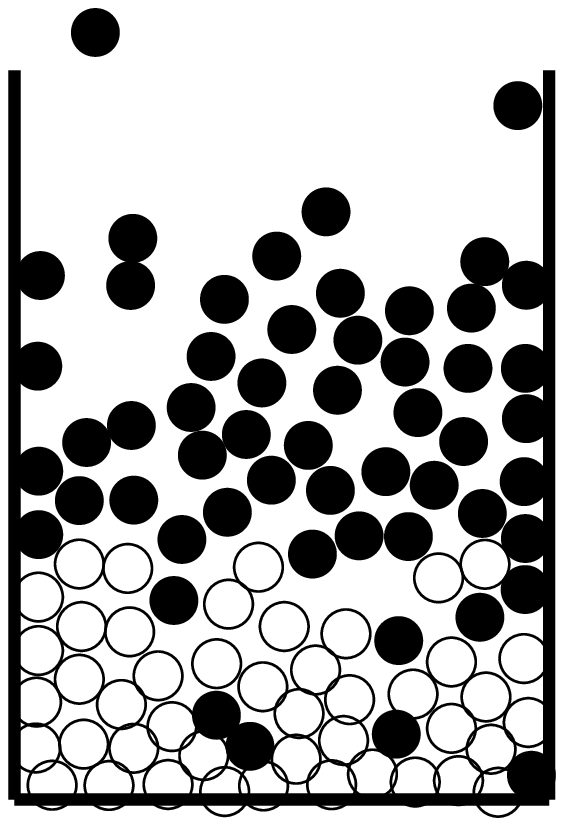}
\includegraphics[scale=0.5]{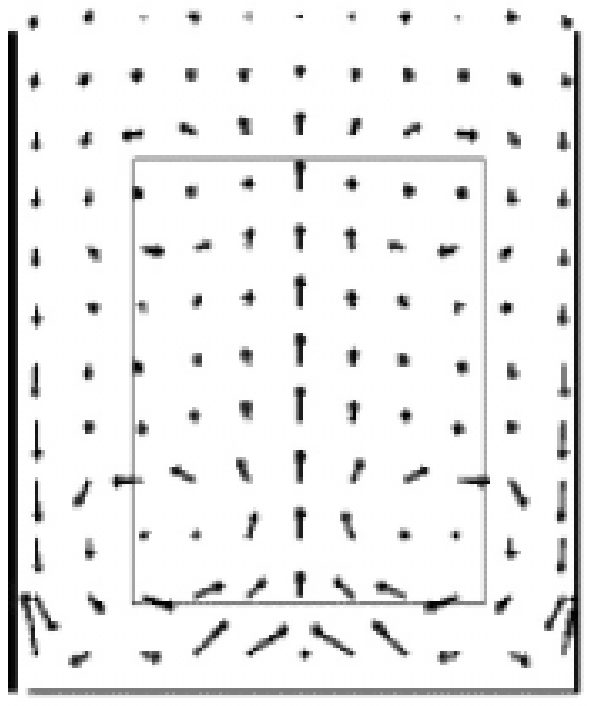}
\caption{ (a) A snapshot of a segregated bed. 
Open circles represent particles with the smaller 
coefficient of restitution.
The two kinds of particles are segregated. 
(b)
Flow vectors of the segregated bed.   
Convective motions are localized at each segregated layer.
The solid rectangle shows the region ($ 3 \leq i \leq 9, 2 \leq j \leq 9$)
where the effective temperature will be computed.}
\label{fig:seg}
\end{center}
\end{figure}

From the numerical point of view, 
a difference from the uniform system occurs only at collisions between
different particles. When this occurs, $k$ and $\eta$ is taken to
be the averaged value of these two. Everything else is the same as the 
previous numerical simulations.
Figure \ref{fig:seg} shows a snapshot and convective flow patterns.
As can be seen easily, two kinds of particles are segregated and
convection occurs in each layer independently.

%%%%%%%%%

\begin{figure}[h]
\begin{center}
\includegraphics[scale=0.8]{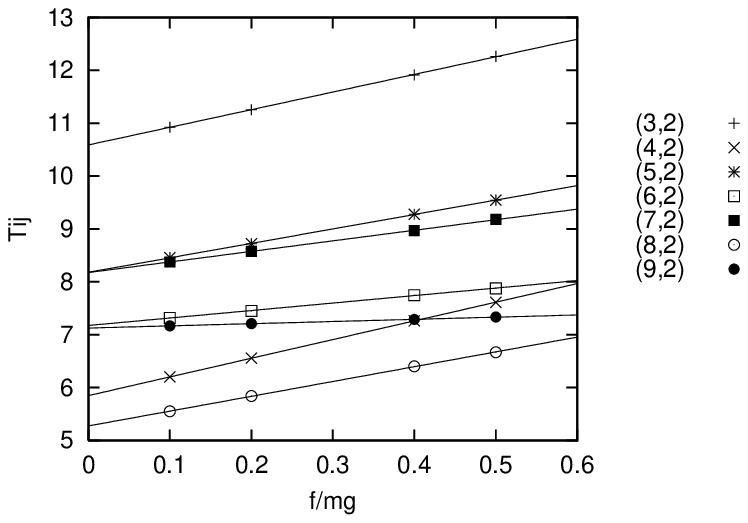}
\caption{The dependence of $ T_{ij}$ upon $f/mg$ at $j=2$.
Solid lines indicate linear extrapolation to $f=0$ using
the least square fit.
The numbers in parentheses indicate the targeted cells as
$(i,j)$.}
\label{fig:T_i2}
\end{center}
\end{figure}

\begin{figure}[h]
\begin{center}
\includegraphics[scale=0.5]{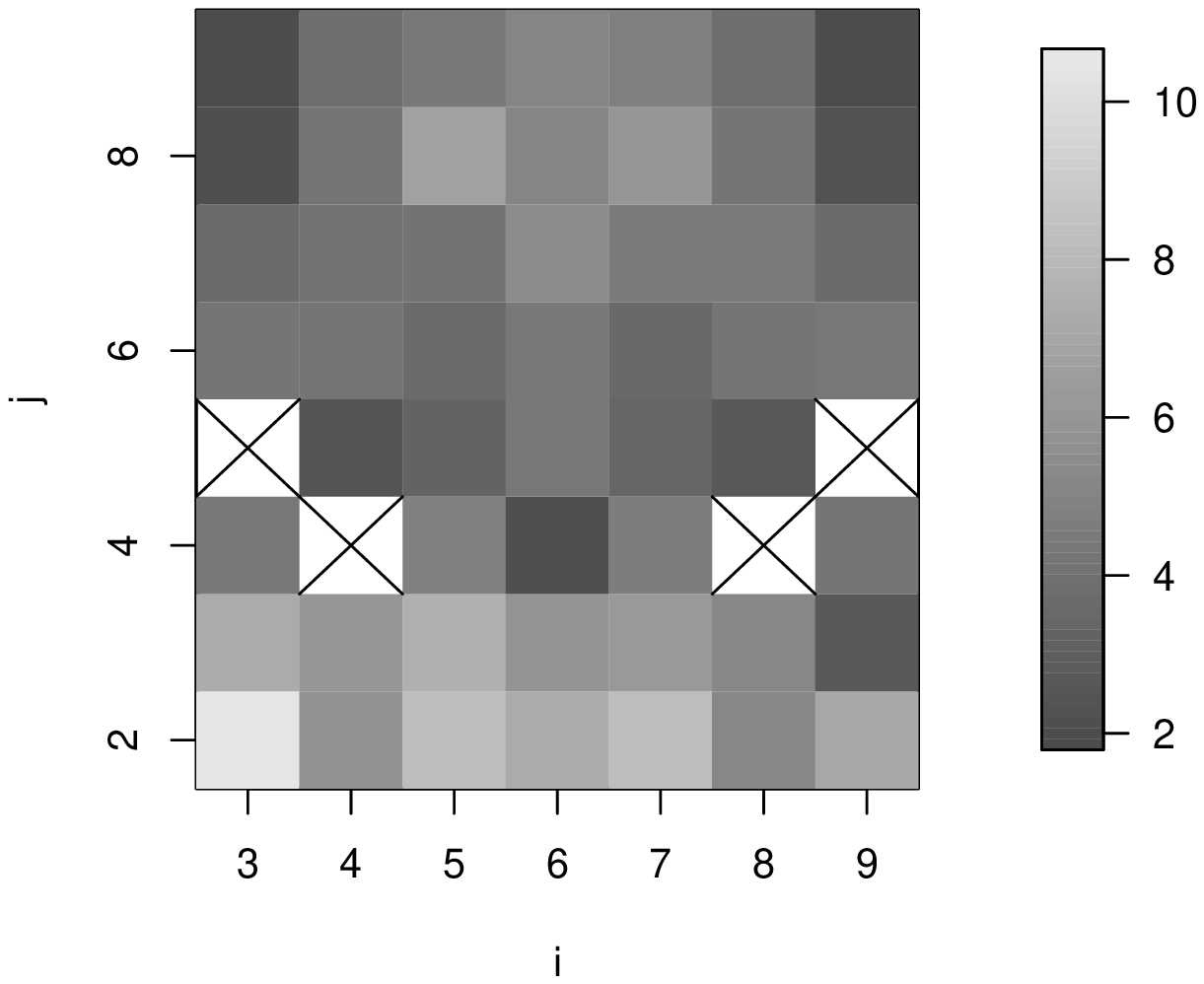}
\caption{Spatial distribution of $ T^{\mbox{eff}}_{ij}$.
Each row corresponds to each horizontal layer of cells.
The whole region corresponds to the rectangular in
Fig. \ref{fig:seg}.
Missing values
( $(i,j) = (3,5),(4,4),(8,4),(9,5))$ are omitted due to the large error.}
\label{fig:T_ij2}
\end{center}
\end{figure}

\begin{figure}[h]
\begin{center}
\includegraphics[scale=0.8]{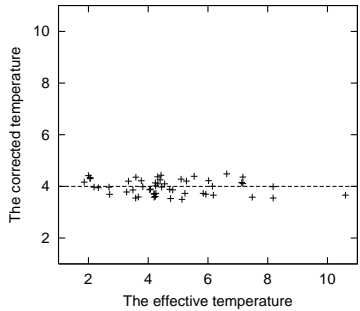}
\caption{The corrected temperature as a function of the effective temperature
$ T^{\mbox{eff}}_{ij}$.
In spite of the large fluctuation of the effective temperature, the
corrected temperature takes almost constant value.
The horizontal line indicates the mean value of the
corrected temperature.}
\label{fig:correct_T2}
\end{center}
\end{figure}

For the calculation of convective flow, 
we do not distinguish different particles.
The effective temperature is measured at each cell as  above.
Figure  \ref{fig:T_i2} shows 
the dependence of temperature at each cell upon $f$.
Again linearity with $f$ is very good and we get extrapolated values
using the least square fits.
Fig \ref{fig:T_ij2}  shows the spatial distribution of temperature.
Clearly spatial distribution is much more scattered
than that in Fig. \ref{fig:T_ij}.
This will be because of inhomogeneity of the system.
However, applying corrections to this as done in the previous section,
we get almost constant values of the effective temperature 
(Fig. \ref{fig:correct_T2}),
although they are a little  more scattered than in the uniform case.
Thus it suggests that the thermal equilibrium state is not destroyed
even if the system is not uniform.
Even when the drastic phenomena like segregation occur,
the thermal equilibrium state can still be maintained.

%\hspace{2.5cm}  \fbox{Fig. \ref{fig:seg}}
%\fbox{Fig. \ref{fig:T_i2}}
%\fbox{Fig. \ref{fig:T_ij2}}
%\fbox{Fig. \ref{fig:correct_T2}}

\section{Discussion}
\label{sec6}
Although we have found a thermal equilibrium state using
the temperature defined by us, there are still some
uncertain points in this study.
First, the heuristic arguments presented are not substantiated by any
first-principles argumentation and the success of the study is judged based on
the good uniformity of the effective temperature.
Actually our finding that we have obtained a thermal equilibrium
state does not always justify our heuristic argument of the temperature.
However, even if we cannot justify our results completely,
we believe that our findings cannot be accidental.
It was very difficulty to find a phenomenological temperature 
definition which characterizes the thermal equilibrium state
in the dynamical state of  granular matter.
Furthermore, our definition is based on the foregoing studies 
\cite{makse,D'Anna}, thus it has some
physical bases although
it is not very complete. Our findings can be a start point of
understanding general dynamical properties of granular matter.

Second, the numerical model studied here appears to be slightly
artificial (two-dimensional, relatively few particles, cells of the same size
as the particles, making it impossible to study systems composed of
non-monodisperse particles) and therefore the results obtained with this model
are not as convincing as they might have been if a different, less restricted
model had been employed. The reason why we had to employ such a model
is mainly because of the computational limitation.
In spite of the apparent  easiness of our simulations,
it is extremely time consuming. 
We need very long time simulations in order to suppress the
fluctuations. This is because we need the absolute values of
fluctuations, not mean values. 
Compared with getting mean values,
measuring fluctuations quantitatively is much more time consuming.
Furthermore, we have to repeat numerical simulations because
we need extrapolations and 
the number of points which we can compute at the same time 
is only a few, as noted  above.
Of course, it is much more suitable to reproduce
these results by more realistic numerical simulations, but
it is beyond the scope of this paper.
Validating them using more suitable systems will be another
future issue.

\section{Summary and Conclusion}
\label{sec7}

In this study of a granular system, 
we have introduced heuristic procedures to measure
the effective temperature (corrected) proposed by Makse
and Kurchan \cite{makse}.
In the vibrated bed,
thermal equilibrium is maintained even if spatial uniformity
is kinetically  destroyed by  convective motion.
Also it is still maintained even when segregation occurs in the system.
Since the effective temperature is introduced only phenomenologically,
it is important to understand this phenomenon  by  statistical mechanics.
However, since clearly energy is not conserved,  how to
construct statistical mechanics in  granular matter is unclear.
Although we do not have any clear pictures at the moment, 
recent findings \cite{Ojha}
strongly suggest that statistical mechanics is apparently valid even for
the single particle system without collisions.
Probably, there is some robustness in the apparent
thermal equilibrium system of the dynamics of  granular matter.

\end{document}